\documentclass[12pt,preprint]{aastex}

\shorttitle{STRUCTURE MAPS OF CFA SEYFERTS}
\shortauthors{POGGE \& MARTINI}

\begin{document}

\title{{\it Hubble Space Telescope} Imaging of the Circumnuclear 
Environments of the CfA Seyfert Galaxies: Nuclear Spirals and
Fueling\footnote{Based on observations made with the NASA/ESA Hubble
Space Telescope, obtained from the data archive at the Space Telescope
Science Institute.  STScI is operated by the Association of Universities
for Research in Astronomy, Inc., under the NASA contract NAS5-26555.}}

\author{Richard W. Pogge \& Paul Martini\altaffilmark{1}}
\affil{Department of Astronomy, The Ohio State University}
\affil{140 West 18$^{th}$ Avenue, Columbus, Ohio 43210-1173\\
pogge@astronomy.ohio-state.edu, martini@ociw.edu}

\altaffiltext{1}{Current Address: Carnegie Observatories,
813 Santa Barbara St., Pasadena, CA 91101}


\begin{abstract}

We present archival {\it Hubble Space Telescope} images of the nuclear
regions of 43 of the 46 Seyfert galaxies found in the volume-limited,
spectroscopically complete CfA Redshift Survey sample.  Using an
improved method of image contrast enhancement, we create detailed
high-quality ``structure maps'' that allow us to study the distributions
of dust, star clusters, and emission-line gas in the circumnuclear
regions (100-1000 pc scales) and in the associated host galaxy.
Essentially all of these Seyfert galaxies have circumnuclear dust
structures with morphologies ranging from grand-design two-armed spirals
to chaotic dusty disks.  In most Seyferts there is a clear physical
connection between the nuclear dust spirals on hundreds of parsec scales
and large-scale bars and spiral arms in the host galaxies proper.  These
connections are particularly striking in the interacting and barred
galaxies.  Such structures are predicted by numerical simulations of gas
flows in barred and interacting galaxies, and may be related to the fueling
of AGN by matter inflow from the host galaxy disks.  We see no
significant differences in the circumnuclear dust morphologies of
Seyfert 1s and 2s, and very few Seyfert 2 nuclei are obscured by
large-scale dust structures in the host galaxies.  If Seyfert 2s are
obscured Seyfert 1s, then the obscuration must occur on smaller scales
than those probed by HST.

\end{abstract}

\noindent{Accepted for publication in the ApJ, v569 (2002 Apr 20)}




\section{Introduction} \label{sec:intro}

There is now a large and compelling body of observational evidence that
suggests that most, if not all, galaxies contain supermassive black
holes at their centers \citep[e.g.,][]{richstone98}.  However, active
galactic nuclei (AGN) are only found in a small minority of all galaxies
in the local universe \citep[e.g.,][]{huchra92,ho97}.  What is it that
makes some galaxies AGN but most others quiescent?  One line of inquiry
is to ask if the differences are to be found in their circumnuclear
environments.  In particular, is the difference simply a matter of
whether or not the central black hole is being provided with interstellar
gas to fuel the nuclear activity?

The problem of providing fuel to an AGN from the vast reservoirs of
interstellar gas found in the disks of spiral galaxies is how to remove
the angular momentum from the gas so it can fall into the nucleus.  The
two classical mechanisms that are invoked are interactions
\citep{toomre72} and bars \citep{schwartz81}, including nuclear bars
\citep{shlosman89,pfenniger90}.  There is an extensive literature devoted
to demonstrating that both are theoretically viable mechanisms for
fueling AGN
\citep{hernquist89,barnes91,athanassoula92,friedli93,piner95,hernquist95}.
However, neither interactions nor bars of either type are sufficiently
common among AGN compared to non-AGN galaxies to be the fueling
mechanism in all cases
\citep{adams77,petrosian82,keel85,fwstocke88,kotilainen92,mcleod95,keel96,
alonso96,mulchaey97,derobertis98,regan99,knapen00,schmitt01,laine01}.

Previous investigations have used {\it HST} to study the circumnuclear
environments of Seyfert galaxies and search for differences between the
Seyferts with and without broad-line components. \citet{nelson96}
obtained pre-COSTAR imaging of a large sample of Seyfert and non-Seyfert
Markarian galaxies to look for differences in the nuclear structure of
these galaxies at higher angular resolution than is possible with
ground-based imaging. They discovered that the nuclei of broad-line
Seyfert $1-1.5$ galaxies are dominated by strong point sources. In
contrast, Seyfert $2$ galaxies and other Markarian galaxies that lack
broad-line regions contained weak or no strong nuclear source
superimposed on the underlying galaxy's surface brightness profile. This
result is further borne out in the extensive {\it HST} snapshot program
of \citet{malkan98}. They also invariably find more strong central point
sources in Seyfert 1 galaxies than in the Seyfert 2s.  As this survey
was carried out with the unaberrated WFPC2 PC camera, these
investigators were also able to look for differences in the nuclear
environments of these galaxies. They found Seyfert 2 galaxies were more
likely to possess dusty nuclear environments than Seyfert 1 galaxies,
lending support to unified models, which propose obscuration of the
broad-line region in Seyfert 2s by dust. Their observations are evidence
for the presence of dust on large scales in the nuclear region, not in a
torus immediately outside the broad-line region.

Visible--near-infrared color maps obtained with the {\it Hubble Space
Telescope} ({\it HST}) have shown that the circumnuclear ($\sim 100 -
1000$ pc) regions of a large number of low-luminosity AGN contain
nuclear spiral dust lanes. These spirals are distinct from the spiral
arms on kpc scales in the main galaxy disk. $V-H$ color maps of these
galaxies show that these `nuclear spirals' extend from 100's of pc
scales into the unresolved nucleus \citep{quillen99,regan99}.
Theoretical models for the formation of nuclear spiral structure suggest
that it is dynamically distinct from the main disk spiral arms
\citep{bertin89,elmegreen92}. \citet{martini99} showed that nuclear spirals 
in AGN reside in nonself-gravitating disks and are therefore likely due to
shocks in nuclear gaseous disks. They postulated that as shocks can
dissipate energy and angular momentum, these nuclear spirals may be the
signature of the fueling mechanism in these galaxies.

Nuclear spirals in AGN have generally been seen mostly in Seyfert 2s,
although this could easily be a selection effect: the samples of
\citet{quillen99} and \citet{regan99} were mostly comprised of Seyfert
2s. Also, \citet{martini99} only observed Seyfert 2s with {\it NICMOS}
as they have fainter nuclear PSFs, and thus the circumnuclear
environments of Seyfert 2s are easier to study with {\it HST} than the
circumnuclear environments of Seyfert 1s. The question remains, however,
if Seyfert 1s and 2s contain nuclear spirals with the same relative
frequency; that is, nearly 100\% as seen in Seyfert 2s.
As we only have near-infrared {\it NICMOS} imaging of the Seyfert 2s in
the CfA sample, we need to use an alternate technique to look for
nuclear spiral structure in these Seyferts.  In \S\ref{sec:procedure} we
discuss our data-processing, and introduce a technique for creating
``structure maps'' in \S\ref{sec:stmaps} that are an excellent surrogate
for color maps for detecting small-scale dust-extinction and
emission-line features present in the visible-band images.  We then use
this technique to compare the circumnuclear environments of the Seyfert
1s and 2s in \S\S\ref{sec:morph} and \ref{sec:sey}, connecting the nuclear 
structures seen to the larger host galaxies in most cases. In 
\S\ref{sec:conc} we present a summary of our results, and discuss the 
implications for the fueling of the active nuclei.


\section{Sample Selection} \label{sec:sample}

A limiting factor in statistical studies of Seyfert galaxies is the
notorious difficulty of identifying a homogeneous sample free of
crippling selection biases.  Since Seyferts are identified primarily by
their line spectra, one way to be absolutely certain of completeness is
to take a spectrum of {\it every} galaxy down to some sensible magnitude
limit.  At present, the largest spectroscopically complete host-galaxy
selected sample of Seyfert galaxies is the \citet{huchra92} sample
derived from the CfA Redshift Survey \citep{huchra83}.  While not
ideally homogeneous, its selection criteria are nonetheless very well
understood.  The CfA survey obtained optical spectra of a complete
sample of 2399 galaxies down to a limiting photographic magnitude of
$m_{Zw}\le 14.5$ in fields limited to $\delta\geq0\arcdeg$ and
$b_{II}\geq40\arcdeg$, and $\delta\geq-2.5\arcdeg$ and
$b_{II}\leq-30\arcdeg$.  This spectroscopically complete sample of
Seyfert galaxies is reasonably large by AGN standards (46 galaxies), and
relatively free of most of the usual selection biases, especially biases
against reddened AGNs that are inherent in traditional UV-excess
surveys.

A further advantage of the CfA sample is that \citet{martel93} have
obtained high signal-to-noise ratio (S/N) spectra of the nuclei of
galaxies identified as having narrow lines (primarily Seyfert 2s).  This
provides us with a definitive set of spectral classifications for all of
the CfA Seyfert galaxies with uniform depth.  This important property is
often overlooked in compiling samples of Seyfert galaxies.  Much of the
classification material in the literature is of mixed quality,
especially for the intermediate types 1.5, 1.8, and 1.9
\citep{osterbrock81}, where the broad-line component can be relatively
weak.  Without a carefully and homogeneously acquired set of
classification spectra, a sample will be biased against the detection of
nuclei with weak broad-lines because such galaxies will be misclassified
as Seyfert 2s.  This is a potential problem with large samples selected
out of the general catalogs without regards to the quality of the
available classification material, especially if such samples are going
to be used to examine the statistical properties of Seyferts with and
without broad-line regions in direct spectra.  A recent criticism of the
CfA Seyfert sample is that it missed many low-luminosity AGN with weaker
emission lines due to the fact that the spectra were obtained for the
purpose of redshift estimation \citep{ho01}.  Our comparison of the
circumnuclear regions of Seyfert 1s and 2s is thus based primarily upon
objects with relatively bright emission lines. In this paper we will
refer to the Seyfert 1 and 1.5s collectively as the ``Seyfert 1s'' and
the Seyfert 1.8, 1.9, and 2s collectively as the ``Seyfert 2s.''  This
division reflects the fact that most of the 1.8s and 1.9s were at
one time classified as Seyfert 2s, and is thus a division between
``strong broad lines'' and ``weak broad lines'' rather than an arbitrary
division among what is arguably a continuum of types.  Lumping the 1.8s
and 1.9s in with the ``Seyfert 1s'' would only have the effect of giving
us a much smaller number of Seyfert 2s, but it does not otherwise affect
our results significantly as will become apparent.

The final virtue of this sample is that it is volume limited.  For all
Seyfert types, $V/V_m$ tests \citep{schmidt68} show that both Seyfert 1s
and 2s have $\langle{V/V_m}\rangle\approx0.5$ to within the
uncertainties \citep{huchra92}. \citet{martel93} repeated this analysis
using different combinations of the various intermediate types, showing
that this result is unaffected by how the Seyferts are subdivided into
different types.  The only systematic effect found thus far is that
luminosity functions of the CfA Seyferts as a function of Seyfert type
\citep{huchra92,martel93} show that the integrated (host+nucleus)
luminosities of the Seyfert 1s are on average about 1 magnitude brighter
than the Seyfert 2s.  The results of \citet{nelson96} suggest that this
is primarily due to the systematically brighter Seyfert 1 nuclei.

Out of the 46 CfA Seyferts in our sample 43 ($\sim 94$\%) have been
imaged in broad bandpasses by the {\it Hubble Space Telescope} ({\it
HST}) with the Wide Field and Planetary Camera 2 (WFPC2).  In general,
the archival PC1 camera imaging in broad-band filters (most of which are
snapshots) constitute a relatively uniform set of images with sufficient
depth and spatial resolution to reveal detailed structures at
subarcsecond scales in the nuclear regions.  Most of these structures
are expected to be due to stars and dust in the host galaxy.  Extended
emission-line regions can also be seen in these broad-band images, if
they are high surface-brightness regions composed of essentially
unresolved knots of emission, even though the bandpasses are broader
than used in narrow-band imaging work with {\it HST}
\citep[e.g.,][]{pogge97}.  These images will permit us to map out
the immediate (few 100~pc) environment of nearby Seyfert nuclei in a
well-defined, spectroscopically complete sample of objects.


\section{Data Collection and Processing} \label{sec:procedure}

\subsection{The WFPC2 Imaging Sample}

We have searched the HST Archives for all publicly available WFPC2
images of the CfA Seyfert galaxies listed in Table~1 of
\citet{huchra92}.  We include Mrk\,471 which was accidentally omitted
from Huchra \& Burg's Table~1 (but included in their analysis), and
exclude the starburst galaxy Mrk\,789 \citep[][]{martel93,dahari88}.
For this work we have selected only those galaxies classified as
Seyferts and neglected Liners, X-ray galaxies and QSOs.  The Seyfert
classifications are taken from the classifications given by
\citet{martel93}.  The final sample of 46 galaxies, our
adopted classifications, and the archival imaging parameters (exposure
time, filter, and GO program ID), are listed in Table~\ref{tbl:cfa}.
Distances to each galaxy in Mpc are listed in Column 7, computed by
transforming the heliocentric radial velocities given by
\citet{huchra92} into the rest frame of the Local Group following
\citet{yahil77} and adopting $H_0=75$\,km\,sec$^{-1}$\,Mpc$^{-1}$.
Column 8 gives the projected scale in parsecs corresponding to 1\arcsec\
on the sky at the distance of the galaxy.  Archival {\it HST/WFPC2}
images of suitable quality for our purposes are available for 43 of the
46 CfA Seyferts.  There is no archival WFPC2 imaging for three of the
galaxies among the 46: Mrk\,841, NGC\,3080, and 1335+39.  The absence of
these galaxies from our sample does not significantly affect our
results.

Most of the galaxies (35 out of 43) have 500-second F606W (wide-V) band
WFPC2 Planetary Camera (PC) images obtained as part of a large {\it HST}
Cycle 5 snapshot survey of AGNs (GO5479, PI: Malkan).  Images of the
remaining galaxies are available from archival data of individual
observations made by a variety of programs (Table~\ref{tbl:cfa}, Column
6).  For these galaxies, we used the F547M filter or F814W filter
(I-band) images if F606W images were unavailable.

In the case of NGC\,1068, all of the PC detector pixels in the central
2--3\arcsec\ of the nuclear region were completely saturated in the
500-second F606W snapshot image from GO5479. For this reason we also use
the archival F547M images to recover information about the nucleus.
These data consist of a pair of short integrations (140 and 300-seconds)
that are not saturated.  We will use these to probe structures closer
than $\sim$1\arcsec, and use the deeper F606W snapshot for the outer
regions.

\subsection{Post-Pipeline Image Processing}

All of the WFPC2 images used in this study were already processed by the
standard STScI reduction pipeline \citep[described by][]{biretta96}.
After inspection of the post-pipeline images to verify that they were of
acceptable quality for this study, the only additional processing steps
required were removal of cosmic rays, and treatment of
saturation-related artifacts in the nuclei (destreaking).  All of the
galaxies selected had the active nucleus roughly centered in the PC
detector, so we did not process any of the WF camera frames.

The majority of the images (35 of the 43) are snapshots acquired in a
single exposure.  In these cases, we used the {\sc cosmicrays} task in
the NOAO {\sc IRAF}\footnote{{\sc IRAF} is distributed by the National
Optical Astronomy Observatories, which are operated by the Association
of Universities for Research in Astronomy, Inc., under a cooperative
agreement with the National Science Foundation.} package to
statistically remove most of the cosmic ray events.  Cosmic ray events
were defined as pixels deviating by more than 5$\sigma$ above the local
mean computed in a $7\times7$ pixel window, and were replaced by the
average of the four neighboring pixels.  Any residual cosmic ray events
were hand-cleaned from the final images with an interactive median
filtering tool in the {\sc XVista}\footnote{{\sc XVista} is the direct
lineal descendant of Lick Observatory Vista, and is maintained in the
public domain by a cadre of former Lick graduate students as a service
to the community at {\tt http://ganymede.nmsu.edu/holtz/xvista}}
package.

For those galaxies with pairs of images of the same (or similar)
integration time, we combined the images using a statistical
differencing technique implemented as a {\sc XVista} procedure script.
This latter technique works as follows.  The difference image formed by
subtracting one image in the pair from the other consists primarily of
positive and negative cosmic ray hits, as the galaxy, foreground stars,
and background all cancel to within the uncertainties.  All pixels
within $\pm5\sigma$ of the mean residual background level on the
difference image are then set to zero (tagging them as unaffected by
cosmic rays).  A pair of cosmic-ray template images is then formed by
separating the remaining positive and negative pixels.  These templates
are subtracted from the original images, and the two cosmic-ray
subtracted images are added together to form the final galaxy image.

A number of the images have the nuclei over-exposed and saturated. We
judged saturation by consulting the Data Quality File (DQF) that
accompanies each WFPC2 image \citep[][]{biretta96}.  Two forms of
saturation were encountered: pixels at or above the maximum A/D
converter level (ADC saturation), and pixels that are so saturated they
overfill the wells and bleed charge into adjacent pixels in their column
(full-well saturation or ``bleeding'').  In our descriptions below, a
nucleus was judged to be ``completely saturated'' if (1) it showed bleed
streaks due to full-well saturation, or (2) the central 9 or more pixels
were flagged as saturated in the DQF, corresponding to all pixels within
the FWHM of the PC1 point-spread function in this band. In general, it
takes only one or two pixels to be saturated in the core of the PSF to
obviate deriving any useful quantitative photometry of the nucleus.  For
this reason our analysis will concentrate on morphological
characteristics.  Bleed streaks of between 1 and 5 pixels in width
appear in the most saturated nuclei, and would complicate our structure
mapping analysis described in the following section if left alone.  We
destreaked these images via a simple method that successively
interpolated across the streaks from the inside out until they are
removed.  No significant artifacts from the destreaking process are seen
in our final structure maps.

\section{Structure Maps} \label{sec:stmaps}

Ideally, we would like to map the circumnuclear dust in these galaxies
by creating photometric color maps from pairs of images taken at widely
separated wavelengths, e.g., by making $(V-H)$ color maps as we did in
our previous study of Seyfert 2s \citep{martini99}.  For more than half
of our sample, however, corresponding NICMOS images are not available,
especially among the Seyfert 1s.  Even where NICMOS imaging is
available, the small field of view of NICMOS restricts the maps to the
central 5--10\arcsec\ or so \citep{martini99,regan99}.  We therefore
need an alternative way to recover this information in the absence of
wide-field IR images.

In $(V-H)$ color maps like those we presented in \citet{martini99}, the
regions of strong red or blue color that appear are primarily due to
structures that are strongest in the V-band images.  The H-band images,
by contrast, are very smooth and appear essentially featureless in all
but the most unusual cases.  In creating a $(V-H)$ color map we are in
effect using the H-band image to suppress the underlying starlight
distribution in the V-band image.  The dust and emission structures are
certainly visible in the V-band images, but their contrast is diminished
by the bright stellar background, especially in the innermost regions.
The contrast is further reduced in the Seyfert 1s by the strong,
unresolved nuclear point-source, especially since the core of the PSF is
usually completely saturated in these images, and the surrounding
1--2\arcsec\ are contaminated by light from the wings of the WFPC2
point-spread function \citep[e.g.,][]{biretta96}.

One way to enhance such fine structural features in single-filter images
is to fit the smooth galaxian starlight profile with elliptical
isophotes and then subtract or divide the model fit from the images.
The fine structure emerges as fit residuals.  This is often done to
search for dust and other structures in ground-based and {\it HST}
images of elliptical galaxies
\citep[e.g.,][]{sparks85,mcnamara93,vandokkum95,mcnamara96,koekemoer99}.
For the great majority of our galaxies, however, this technique is
impractical because the nuclear-region isophotes are generally not
elliptical and isophote fitting fails completely, producing residuals
that are more a reflection of the fit's failure to converge rather than
real structures.  The presence of a bright, saturated nucleus in many of
our images makes matters worse.

We have instead devised an alternative approach based on Richardson-Lucy
(R-L) image restoration \citep{richardson72,lucy74} applied to {\it HST}
imaging \citep{snyder93} that yields excellent results, as we shall
describe and demonstrate.  The method is as follows (why it works is
described below).  For each cleaned image, we convolve a copy of that
image with a model of the PSF for the PC1 camera appropriate to the
filter bandpass generated using {\sc TinyTim} \citep[][]{kristhook99}.
We adopted an Elliptical galaxy template spectrum for computing the
polychromatic model PSF.  We then divide the original image by the
PSF-smoothed image, and then further convolve this ratio with the
transpose of the model PSF. Mathematically, this procedure produces a
final image $S$ defined as:
\begin{equation}
S = \left[\frac{I}{I \otimes P}\right]\otimes P^{t}
\end{equation}
where $I$ is the original image, $P$ is the model PSF, and $P^{t}$ is
the transpose of the model PSF, $P^{t}(x,y)=P(-x,-y)$, and $\otimes$ is
the convolution operator.  (The use of $P^{t}$ reveals the origins of
this technique in R-L restoration, as shall be described below).  The
procedure described above has been implemented as an {\sc XVista} script
that performs all of the convolutions in the Fourier domain using
Fast-Fourier Transforms.

We call the resulting image $S$ a ``structure map'' because the process
described above has effectively removed most of the larger-scale smooth
light distribution, highlighting unresolved and marginally resolved
structures on the scale of the PSF.  Dusty regions appear dark, while
compact emission-line regions or clusters (knots) of stars appear
bright, much as they would in a color map derived from two images.
These maps recover both dust and emission structures with very high
fidelity, and with fewer artifacts due to mis-matched PSFs in the inner
regions as seen in visible/IR color maps.

Figures~\ref{fig:stmap1} -- \ref{fig:stmap7} and Figure~\ref{fig:n1068}
(NGC\,1068) show the structure maps for the central 30\arcsec\ of each
of the 43 CfA Seyfert galaxies in our sample.  The intensities (black to
white) show the fractional residuals ($\pm10\%$) about the original
pixel-to-pixel intensity.  Dark regions show the locations of dust
obscuration, and bright regions are either locations of enhanced stellar
light (e.g., star formation regions) or emission-line regions (the F606W
filter is wide enough to admit several bright emission lines from high
surface-brightness regions).  Images appear in the same order as listed
in Table~\ref{tbl:cfa}, running from left-to-right and top-to-bottom
across the figure, in order of Seyfert 1 through Seyfert 2.  Within each
Seyfert type, the galaxies appear in alphanumeric order by the name
listed in Column 1 of Table~\ref{tbl:cfa}.  The scale bar in each image
panel indicates 1~kpc projected size at the adopted distance of the
galaxy (Table~\ref{tbl:cfa}, Col 8).  Figure~\ref{fig:n1068} shows
NGC\,1068 separately as its close proximity and large physical size
allows sampling of particularly fine spatial scales.  The left panel of
Figure~\ref{fig:n1068} shows an F606W filter image with the same
30\arcsec\ field of view used in the other structure maps, while the
right panel shows the central 10\arcsec\ of a short-exposure F547M
filter image.  A detailed description and interpretation of these
structure maps is provided in the next section.  When referring to
images of individual galaxies in the text below, we will give the figure
number and the image's location in the figure the first time it is
cited.  For example, the image of UM\,146 is Fig\,\ref{fig:stmap5}
middle right.

Structure maps of visible-wavelength WFPC2 images are excellent
surrogates for color maps, allowing us to recover a great deal of
information from a single-band image.  To illustrate this,
Figure~\ref{fig:vh} shows (left) a $(V-H)$ color map of the central
5\arcsec\ of NGC\,7674 from \citet{martini99} and (right) the structure
map of the same region derived from the V-band image alone using the
techniques described above.  The structure map shows all of the dust
(dark) and emission (bright) features visible in the $(V-H)$ color map,
but without artifacts due to the mismatch between the WFPC2 and NICMOS
point-spread functions.  The structure maps give us two advantages over
the $(V-H)$ color maps derived from WFPC2 and NICMOS images: we can get
closer to point-like nuclei at full WFPC2 resolution (higher than that
of NICMOS), and we can recover information across most or all of the
$\sim$34\arcsec\, PC1 camera field of view, as compared with the
$\sim$5\arcsec\, field of view of the NIC1 images presented in
\citet{martini99}.  This greater field of view will be crucial for our
analysis as it will allow us to study any associations between nuclear
dust structures and larger-scale structures like bars or disk spiral
arms in the host galaxies proper.

Why does it work?  The structure map defined above is approximately the
``correction image'' that emerges from the second iteration of a R-L
image reconstruction \citep[see][Eqns 5 \& 6]{snyder93,richardson72}.
At the end of each iteration of an R-L reconstruction, the resulting
reconstructed image is convolved with the PSF and compared to the
original image.  A correction image is generated by forming the ratio of
the original image to its PSF-convolved reconstruction which measures
the fractional deviation between the model and the original.  This
information is then used to refine the estimate of the reconstructed
image during the next iteration.  This correction and comparison process
is repeated until the correction image no longer changes significantly,
usually because noise and artifacts, relentlessly propagated through
each iteration, overwhelm whatever residual structures remain to be
reconstructed.  After the second iteration, the corrector frame contains
the unrecovered structures after the reconstruction has recovered only
the first-order, smooth structure in the image (the first iteration in
typical implementations compares the original image to a
``reconstruction'' which contains no structure at all).  Our structure
map is essentially this second-iteration correction image, which is why
it highlights unresolved and marginally resolved high-order structures
present in the images.  Formally, however, the expression defining the
corrector image from the second R-L iteration contains an additional
convolution with the transpose of the PSF in the denominator (i.e.,
$(I\otimes P^{t})\otimes P$ instead of $I\otimes P$).  This difference
is minor, as operationally it makes the true second-iteration correction
frame slightly smoother than our structure map.

Like any reconstructive technique, structure maps are exquisitely
sensitive to noise and artifacts in the images, especially single bright
or dead pixels.  If these are not removed, the bright negative or
positive false stars appear in the structure map images.  Other
artifacts are the appearance of negative ``moats'' around very
high-contrast point sources (either in the nuclei or nearby field stars)
and bright spots or smudges near the residuals of cosmic ray events (we
can never completely remove the brightest ion-event trails).  Moats are
a problem for those images with strongly saturated nuclei.  We can
eliminate most moats by truncating the nuclear brightness profiles
before creating the structure map, allowing us to recover information
close to a bright nucleus.  In the very brightest nuclei we still lose
information despite this.  Artifacts due to single bad pixels are
removed manually using the interactive pixel zapper ({\sc tvzap}) in
{\sc XVista} and then iterating until all of the most obvious artifacts
are removed from the subsequent structure maps.

While the structure mapping procedure appears superficially similar to
so-called ``digital unsharp masking''
\citep[e.g.][]{heisler94,walterbos94}, the results are quite different.
To illustrate this, the top two panels of Figure~\ref{fig:unsharp} show
(top left) the central 15-arcseconds of the unprocessed F606W image of
NGC\,3516 (displayed as logarithmic intensity) and (top right) the
structure map generated using the method described above.  Below these
we show a pair of normalized unsharp-mask images generated using (left
bottom) a Gaussian kernel with FWHM=2.2 pixels (the FWHM of the core of
the PSF on the PC1 detector in this filter), and (right bottom) a
$3\times3$ pixel boxcar kernel (the smallest practical boxcar kernel
width).  Each of the three maps are displayed to show the same range of
dark and bright features.  The structure maps created by the method
described above are superior in all respects.  No choice of Gaussian,
Boxcar, or other smoothing kernels (e.g., Hanning, Cosine bell, etc.)
could be found that produced results of comparable quality.  In some
sense this should not be surprising, as what passes for unsharp masking
in the literature is all too often the result of an arbitrary choice of
smoothing kernel.  Our method is formally rooted in reconstructive
analysis of the images and uses the image PSF which embodies detailed
information about how the original images are formed, including the
power in the broad wings surrounding the diffraction-limited core.


\section{Morphology} \label{sec:morph}

\subsection{Circumnuclear Structure} \label{sec:circumnuc}

The circumnuclear regions of all of the Seyfert galaxies in our sample,
with exceptions noted below, have dust and emission regions in the inner
kiloparsec.  It is immediately evident from the structure maps that the
distribution of circumnuclear dust in most of these galaxies takes the
form of nuclear dust spirals.  By analogy with the spirals arms of
galaxy disks, we can characterize them as either ``grand design''
(clearly delineated and symmetric two-arm spirals) or ``flocculent''
(spiral-form, but broken into many arms).  Our structure mapping
technique has successfully recovered all of the spiral dust lanes
previously reported in the CfA Seyfert 2s \citep[][]{martini99},
although we can now trace these structures to larger radii than in the
limited field-of-view NICMOS images.

The nuclear dust spirals in most of the galaxies in this sample appear
to connect to larger scale dust lanes to the signal-to-noise limit of
the images.  Some of these dust lanes form nearly contiguous arms that
can be followed for over a full rotation and extend many kiloparsecs in
length, such as those in NGC\,6104 (Fig\,\ref{fig:stmap3} bottom left),
and Mrk\,744 (Fig\,\ref{fig:stmap4} middle left).  The fact that some of
these dust lanes extend for kiloparsecs suggest they are relatively
long-lived and trace the infall of cold, dense material from the host
galaxy disk.  A notable feature of some nuclear spirals is that they are
the inner extension of large-scale, often ``straight'' dust lanes along
the edges of large-scale stellar bars in the host galaxies.
Particularly striking examples of this are Mrk\,766
(Fig\,\ref{fig:stmap1} bottom left), NGC\,5940 (Fig\,\ref{fig:stmap2}
top right), A0048+29 (Fig\,\ref{fig:stmap2} middle right), Mrk\,471
(Fig\,\ref{fig:stmap4} top right), NGC\,5674 (Fig\,\ref{fig:stmap5}
middle left), NGC\,5347 (Fig\,\ref{fig:stmap7} top left), NGC\,5695
(Fig\,\ref{fig:stmap7} top right), and NGC\,7674 (Fig\,\ref{fig:stmap7}
middle right).  This dust morphology is expected from hydrodynamic
simulations of the flow of interstellar gas under the influence of
large-scale stellar bar potentials \citep{athanassoula92}, and is seen
in non-active galaxies, especially in later-type barred spirals
\citep[e.g.,][]{quillen95,regan97}.  We will discuss these structures
further in the next section.

While spiral dust lanes are very common, they are not the only dust
distribution seen.  In some galaxies the distribution of dust is
chaotic, having no clear overall pattern.  The best examples of this are
NGC\,3227 (Fig\,\ref{fig:stmap3} top left), Mrk\,266SW
(Fig\,\ref{fig:stmap6} bottom left), and NGC\,5929
(Fig\,\ref{fig:stmap7} middle left).  These three galaxies are all
strongly interacting and we will discuss them further below.  In two
galaxies, NGC\,4235 (Fig\,\ref{fig:stmap2} top left) and NGC\,4388
(Fig\,\ref{fig:stmap6} bottom right), the host galaxy is sufficiently
inclined relative to the line of sight that we cannot discern the
circumnuclear distribution of the dust.  Only a few galaxies show no
dust structures on $<1$kpc in their nuclei.  These are the Seyfert 1s
Mrk\,231 (Fig\,\ref{fig:stmap1} top left) and Mrk\,335
(Fig\,\ref{fig:stmap1} middle left), and the dwarf Seyfert 1 NGC\,4395
(Fig\,\ref{fig:stmap4} bottom left); although Mrk\,231 does show
larger-scale structure).  The first two are among the most distant
objects in our sample and have extremely prominent nuclei, but there are
distant Seyfert 1s which still show structures on small scales despite
their distance (e.g., Mrk\,279 and Mrk\,471).  The mottling in the
structure map of NGC\,4395 is due to being able to resolve giant stars
in this nearby ($D_{75}=3.6$Mpc) galaxy.  We will not discuss NGC\,4395
further. Table~\ref{tbl:prop} gives capsule descriptions of the
circumnuclear morphologies revealed by our structure maps.

NGC\,5252 (Fig\,\ref{fig:stmap5} top left) and Mrk\,270
(Fig\,\ref{fig:stmap5} bottom right) have flocculent nuclear spirals
embedded in inclined disks on scales of 1--2\,kpc.  In both cases we see
only the half of the disk on the near side of the host galaxy relative
to our line of sight; this is the side of the disk that will obscure the
most starlight ``behind'' the disk from our point of view.  The far side
of the dust disk will be less distinct because there is more galaxian
starlight between us and the back of the disk.  Figure~\ref{fig:nucdisk}
shows the disks outlined in the structure maps.

The flocculent nuclear dust disk in NGC\,5252 is the same one found by
\citet{tsvetanov96} after they divided their WFPC2 continuum images by 
a smooth elliptical isophote fit.  The kinematic properties of this
disk, which also appears in emission, has been described by
\citet{morse98}.  In our structure map images, we find that this disk
has an axis ratio of 0.73, implying a disk inclination of $\sim
43\arcdeg$, with the semimajor axis of $\sim 3$\arcsec\ ($\sim 1.3$kpc)
oriented along $PA\approx 106\arcdeg$.  This orientation is almost
exactly perpendicular to the major axis of the host galaxy stellar disk
($PA\approx16\arcdeg$), but it is misaligned by $\sim 30\arcdeg$
relative to the axis of the famous ionization cone in this object
\citep[$PA\approx163\arcdeg$;][]{tt89}.  These axes are drawn in 
Figure~\ref{fig:nucdisk}a for reference.

In Mrk\,270, the dusty nuclear disk has a semimajor axis of $\sim
11$\arcsec\ ($\sim 2$kpc) oriented along $PA\approx103\arcdeg$ and is
inclined by $\sim 60\arcdeg$ (axis ratio of $\sim0.5$).  By comparison,
the disk of Mrk\,270 is nearly face-on (axis ratio of 0.927 reported by
2MASS).  Mrk\,270 has only small-scale extranuclear emission, seen here
as the white filaments between $1-2$\arcsec\ northeast of the nucleus in
Figure~\ref{fig:nucdisk}b.  This emission cannot be characterized as an
ionization cone per se (see \citet{pogge89a} for ground-based
[\ion{O}{3}] emission-line imaging).  It is more likely to be dusty
material in the farside of the nuclear disk illuminated and ionized by
the nucleus; note how the bright filaments of emission appear to merge
smoothly with the darker (dust) filaments in the foreground.

An additional component of diffuse emission, apparently {\it stellar} in
origin, is seen nearly perpendicular to the gas disks in both of these
galaxies.  In Mrk\,270, this is the diffuse dumb-bell shaped feature
aligned along $PA\approx 161\arcdeg$ and extending to a projected radius
of $\sim 2$\arcsec.  This feature is aligned exactly with the major axis
of the nuclear stellar bar seen in H-band NICMOS images
\citep[][]{martini01}, and the outer edge of the dumb-bell coincides
with the ends of the nuclear bar.  In NGC\,5252, the enhanced starlight
is aligned nearly exactly with the major axis of the stellar isophotes
along $PA=16\arcdeg$, but it is less pronounced and lacks the sharp
cutoff in radius seen in Mrk\,270.  Such features are not unusual among
the structure maps in our sample.  In particular, the ridge lines of the
strong stellar bars in Mrk\,766, A0048+29, NGC\,5674, (and present but
less obvious without contrast enhancement in NGC\,5940) appear as excess
diffuse light in our structure maps.  It is not immediately obvious to
us why our structure mapping technique should enhance such features, 
as they correspond to structures many times larger than the PSF, but
we note that in these latter cases the features we see in the structure
map are clearly present in the unenhanced raw images, and not artifacts
of the enhancement process.

Emission regions appear in the structure maps as bright, often
filamentary structures on many scales.  The larger-scale emission
features, like those seen in Mrk\,573 (Fig\,\ref{fig:stmap6} top right)
and NGC\,3516 (Fig\,\ref{fig:stmap3} top right) are previously known
from ground-based and space-based imaging (cf. \citet{pogge95} or
\citet{ferruit99} for Mrk\,573; \citet{pogge89b} for NGC\,3516).  
Some of these regions are on scales of $\lesssim$1\arcsec, like the
small figure-8 (bicone?) of emission we see in NGC\,7682
(Fig\,\ref{fig:stmap7} bottom left) or the S-shaped region interrupted
by dust in the nucleus of UGC\,6100 (Fig\,\ref{fig:stmap7} bottom
right).

Other ``emission'' features appearing in the structure maps are bright
stellar knots, many appearing along the spiral arms of the host galaxy.
Some of these knots of emission can be quite near the nucleus, signalling at 
least some
circumnuclear star formation in these objects.  The most obvious example
in our sample is the well-known bright starburst ring in NGC\,7469
(Fig\,\ref{fig:stmap2} middle left) \citep{wilson91}, but not all of
these appear to qualify as ``starbursts.''  Interesting cases,
particularly for future study by those interested in circumnuclear star
formation are A0048+29 (a nuclear ring and bar), NGC\,7603
(Fig\,\ref{fig:stmap3} bottom right), Mrk\,334 (Fig\,\ref{fig:stmap4}
top left), Mrk\,744 (Fig\,\ref{fig:stmap4} middle left), and NGC\,7674
(Fig\,\ref{fig:stmap7} middle right).

\subsection{Large-scale Bars}

About 75\% of the host galaxies in the CfA sample are known to be barred
\citep{mcleod95,knapen00}. It is in many of these barred galaxies that we find 
the grand-design nuclear spirals on sub-kiloparsec scales.  Furthermore,
most of the spirals are clearly connected to the host galaxy via dust
lanes running along the bars.  Particular examples are NGC\,4235,
NGC\,5940, Mrk\,817, NGC\,6104, Mrk\,471, UGC\,12138, NGC\,5674,
NGC\,5347, NGC\,5695, and NGC\,7674, all have ``straight'' dust lanes on
large scales in the structure maps in addition to nuclear grand-design
spirals.  These straight dust lanes are expected for gas and dust
compressed in the principal shocks along the leading edges of a bar
\citep[e.g.,][]{athanassoula92}.  All but Mrk\,817, NGC\,6104, and 
NGC\,5695 have bars strong enough to have led to a barred classification
in the RC3 catalog; the RC3 classification for these three galaxies is
given as ``?''.  NGC\,6104, NGC\,5695, and NGC\,7674 were observed to
have bars in the $K-$band by \citet{mcleod95} (noted by the ``BIR'' in
Table~\ref{tbl:prop}). They did not observe Mrk\,817, although
\citet{malkan98} report a distinct bar in Mrk\,817 in their WFPC2 
image (the same image we use to create this structure map), and
\citet{knapen00} list it as barred in their IR imaging.  Mrk\,279
shows some evidence for the long, ``straight'' dust lanes found in many
large-scale bars, but no bar classification is given by either the RC3,
\citet{mcleod95}, or \citet{malkan98}, although \citet{knapen00}
give it a relatively weak bar classification.  Only three strongly
barred galaxies do not contain circumnuclear grand-design spirals:
NGC\,7469, NGC\,3516, and NGC\,5695. NGC\,7469 and NGC\,5695 may contain
them, although the circumnuclear region is difficult to resolve well in
these two objects, while NGC\,3516 clearly has a multi-arm spiral.

Nuclear spiral structure in the gas (and dust) on sub-kiloparsec scales
is predicted by hydrodynamical models of gas flow in the central regions
of barred galaxies.  In many cases these spirals were barely resolved by
the numerical simulations in the inner kiloparsec due to a decreasing
number of SPH particles
\citep[e.g.,][]{athanassoula92,englmaier97,patsis00}.  However, recent
high-resolution hydrodynamical simulations
\citep{englmaier00,maciejewski01} show the formation of structures
remarkably similar to what we see here in single-barred galaxies on
comparable (few hundred parsec) scales. The structures are spiral shocks
in gas with relatively high sound speeds, leading to gas flows directly
into the nucleus (or at least down to the smallest scales resolved by
their models).  At low sound speeds, they find that the spiral shocks
are interrupted by the Inner Lindblad Resonance and the inflows
terminate on a nuclear ring.  Only two of our galaxies have nuclear
rings: NGC\,7469 (Figure~\ref{fig:stmap2}, middle left) and A0048+29
(Figure~\ref{fig:stmap2}, middle right).

Interestingly, none of the unbarred galaxies have grand-design nuclear
spirals.  The nuclear spiral structure in these systems is often
multi-arm and ``flocculent'' in form.  The morphology of the nuclear
spirals in the Seyfert 2s were previously described in
\citet{martini99}.  The Seyfert 1s shown here exhibit similar types of
structures. Particularly striking and well-resolved nuclear spirals are
apparent in NGC\,4051, NGC\,3227, NGC\,7603, and Mrk\,993. NGC\,4051
shows some evidence for a grand-design, nuclear spiral on roughly
kiloparsec scales, but on smaller (several hundred parsec) scales the
structure breaks up into multi-arm spirals.  The spiral structure in
NGC\,3227 is very irregular and the individual components or armlets can
rarely be traced for more than 60 degrees.  The spirals in NGC\,7603 and
Mrk\,993 are much more regular, but appear to have more than two arms.
Flocculent spiral structures are found in both barred and unbarred
galaxies and could be due to acoustic, or pressure-driven, instabilities
in the circumnuclear gaseous disks \citep{elmegreen98,montenegro99}.

\subsection{Interacting Galaxies}

There are three strongly interacting galaxies in this sample: NGC\,3227,
Mrk\,266SW, and NGC\,5929.  These galaxies have very chaotic dust
structures from large scales down to less than a kiloparsec, suggestive
of the transport of a significant fraction of the host galaxy's ISM into
the central regions.  Such massive transport is a feature of
hydrodynamical simulations of interacting galaxies
\citep[e.g.][]{barnes91,barnes92,mihos96,kennicutt98}.  By contrast,
other clearly interacting but decidedly less morphologically disturbed
galaxies in our sample, specifically Mrk\,744 and NGC\,1144, exhibit a
relatively ordered and spiral dust morphology down to less than
kiloparsec scales.  However, both still have very dusty nuclear regions
compared to other galaxies in the sample.

Bar modes which might drive gas inflow are expected in minor merger
systems, although only after about 1\,Gyr has elapsed since the merger
between the host and satellite galaxy \citep{walker96}.  The nearly
complete dusty disorder we see in the nuclear regions of the most
strongly interacting galaxies suggests that the matter transport has
occurred on less than a dynamical timescale, much faster than the
typical slow and steady infall due to a bar potential. These interacting
galaxies clearly have sufficient matter inflow to fuel nuclear activity
on the scales probed by these {\it HST} observations.


\section{Seyfert 1s vs. Seyfert 2s} \label{sec:sey}

At first sight, the differences between Seyfert 1s and 2s appears
striking, as nearly all of the Seyfert 1s and 1.5s have bright,
saturated star-like nuclei showing multiple Airy rings and diffraction
spikes, whereas the Seyfert 1.8s, 1.9s, and 2s have fainter, unsaturated
nuclei.  This strong nuclear brightness contrast was seen by
\citet{nelson96} in WFPC-1 images of Seyferts, and confirmed by
\citet{malkan98} using a larger WFPC2 imaging survey from which many of
our images were derived.

As we shall see, however, the circumnuclear regions do not show such
systematic differences between the two types: only the very bright
nucleus gives any hint as to the spectral type of the nucleus.  A major
goal of our study is to determine if there are any differences in the
circumnuclear morphologies of these Seyfert 1s and 2s. We inspected each
galaxy for evidence of circumnuclear features such as nuclear spirals,
bars, and rings, in addition to dust lanes consistent with a large-scale
bar feeding gas and dust into this region.  Table~\ref{tbl:prop} lists
the galaxies from Table~\ref{tbl:cfa} and provides their morphological
type, angular size ($D_{25}$), and axis ratio ($R_{25}$) from the RC3,
along with our morphological classifications.

To examine them in detail and take into account the range of distances
and sizes of the CfA sample, we first considered a distance-limited
subset restricted to galaxies within 100 Mpc in which we can easily
resolve structure in the central kiloparsec; more distant galaxies are
denoted by a ``D'' in Column\,8 of Table~\ref{tbl:prop}.  In addition,
nearly edge-on ($R_{25} > 0.3$) and strongly interacting galaxies were
excluded.  This leaves us with seven Seyfert 1s and fourteen Seyfert 2s.
All of these galaxies, with the exception of NGC\,4395 (which is
exceptionally close and small) show evidence for nuclear spiral
structure on sub-kiloparsec scales with no obvious differences in
nuclear dust morphology between the Seyfert types.  We also formed a
size-limited subset in which we took into account the variation in
actual physical size of the galaxies. To do this we computed the
projected physical size in kiloparsecs of $D_{25}$ at the distance of
the galaxy (Column 6 of Table~\ref{tbl:prop}) and then the fraction of
this total size contained within the 30\arcsec\ cutouts from the PC
camera (Column 7 of Table~\ref{tbl:prop}). The size-limited subset only
includes sufficiently large Seyferts, defined here as those with less
than 50\% of the galaxy within the $30''$ images; smaller galaxies are
labeled with an ``S'' in Table~\ref{tbl:prop}.  This subset contains
nine Seyfert~1s and sixteen Seyfert~2s.  {\em All of these galaxies,
with the exception again of NGC\,4395, also contain nuclear spiral
structure independent of Seyfert type.}  While the choice of $100$ Mpc
and 50\% for these distance-limited and size-limited subsets is somewhat
arbitrary, they do effectively remove the smallest and most distant
galaxies in which we would not be able to clearly resolve circumnuclear
structure, regardless of their Seyfert type.

Looking at our entire sample, what we do not see is a tendency for
Seyfert 2 nuclei to be more frequently crossed by host galaxy dust lanes
as found by \citet{malkan98} for a larger, but arguably more
heterogeneous sample of Seyferts. This is likely due to the different
Hubble type distributions of our respective samples. The median Seyfert
1 host studied by \citet{malkan98} is earlier than the median Seyfert 2
type, while the Seyfert 1s and 2s we study have a similar Hubble type
distribution. This difference in host galaxy type can explain why
\citet{malkan98} found that Seyfert 2s were dustier than Seyfert 1s,
whereas we do not find a similar trend.  Within the context of the
standard unification hypothesis, \citet{malkan98} invoke the apparent
differences in the circumnuclear environments of Seyfert 1s and 2s to
posit that galactic dust on hundreds of parsec scales could obscure
Seyfert 1s and make them appear as Seyfert 2s. They suggest that this
``galactic dust model'' is a viable alternative to the classical torus
model \citep{antonucci93}, where the obscuration that gives rise to the
Seyfert 1/2 dichotomy lies on parsec scales.

We do, however, have examples of nuclei obscured by host-galaxy dust in
our sample; particularly striking examples are the interacting galaxies
Mrk\,266SW (Figure~\ref{fig:stmap5}, bottom left) and NGC\,5929
(Figure~\ref{fig:stmap7}, middle left), and the edge-on galaxy NGC\,4388
(Figure~\ref{fig:stmap6}, bottom right), where the nuclei are only
visible at IR wavelengths \citep[e.g.,][]{martini01}. These examples
offer qualitative support of the galactic dust model.  The fact that we
find similar circumnuclear structure in Seyfert 1s and 2s at fixed
Hubble type does not refute the galactic dust model as the actual dust
structures responsible for the nuclear obscuration in any given Seyfert
2 are still likely to be unresolved in most cases (with perhaps the few
exceptions noted above) due to the small sizes of molecular clouds.  We
suspect that the root cause of our discrepancy with the results of
\citet{malkan98} reflect differences in the distribution of host galaxy
morphological types between the Seyfert 1s and 2s in the respective
samples.  The different distributions in \citet{malkan98}'s sample may
be due to the fact that most of the galaxies in their sample are
Markarian Seyferts, selected on the basis of UV-excess.  By contrast,
samples like the CfA Seyferts presented here and the Palomar Seyferts
\citep{ho97} have similar Hubble type distributions between Seyfert 1s
and 2s, and these surveys are both based on spectroscopy of large
numbers of galaxies in a flux-limited survey.

\section{Conclusions}\label{sec:conc}

Using a new ``structure mapping'' technique, we have found nuclear
spiral dust structures that are plausibly related to the inflow of
interstellar gas from the host galaxies into the nuclear regions of a
well-defined sample of Seyfert galaxies.  We find the dust morphology
that is expected if interstellar gas is being driven primarily by
large-scale bars, and if interstellar gas is mass-transported inwards by
torques arising from tidal interactions.  There is also circumstantial
evidence for the spiral infall of gas, with greater or lesser degrees of
coherence, in galaxies which show neither stellar bars nor evidence of
tidal interaction.  In most cases the spiral dust structures seen
previously at small scales in more traditional color maps are shown to
be connected to the large-scale properties of the galaxies, although
they are not simply continuations of the large-scale features.  All of
these structures are consistent with the idea that interstellar gas from
the host galaxy may be transported into the nucleus, to varying degrees
of intensity and coherence, via a variety of mechanisms.

The case that these structures are related to fueling the AGN is
primarily phenomenological.  Theoretical models of gas inflow in barred
and interacting galaxies predict structures similar in appearance to
those we see in our images.  However, what we do not know is whether we
can make the final connection between these larger-scale structures and
the central black hole.  That we find these structures in most of the
Seyferts studied here makes this connection plausible, but we cannot yet
make it conclusively.  From an observational perspective, a particularly
difficult challenge is how to establish kinematic evidence of inflow in
cold interstellar material on these scales.  From the theoretical side,
work on what inflow rates are expected, and what observational factors
(either in addition to or in the absence of direct kinematical
measurements) might help inform such estimations on an object-by-object
basis.

It has long been known that bars and interactions occur in inactive
galaxies, and there is no reason to expect that they will not funnel gas
into the inner regions of those galaxies in much the same way.  Further,
nuclear spirals have been observed in a number of normal spiral galaxies
without obvious AGN activity
\citep{phillips96,carollo98,elmegreen98,laine99}.  If nuclear spirals
are signatures of shocks that can dissipate sufficient angular momentum
to fuel a black hole, the question remains why normal galaxies that 
exhibit such structures are not active, particularly as it is now clear
that essentially all galaxies harbor nuclear black holes, with masses
closely correlated with their host galaxy properties
\citep[e.g.][]{richstone98}.  This suggests that the transport of gas
into the central ($<100$pc) region is a necessary, but not unique
requirement of nuclear activity.

The near-ubiquity of coherent, circumnuclear dust structures suggestive
of shocks and matter inflow supports the conjecture that they are
responsible for fueling their AGN. The first step to test this is a
study of a well-defined control sample of quiescent spiral galaxies of
similar morphological type observed and analyzed in similar ways to the
Seyferts.  We are currently engaged in an on-going HST snapshot program
to address this question of the relative frequency and strength of
nuclear dust spirals in normal galaxies compared to AGN.  More data will
be forthcoming.


\acknowledgements

We wish to thank Drs. P. Osmer, B. Peterson, A. Gould, D. Weinberg,
S. Mathur, L. Ho, M. Regan, J. Mulchaey, D. Maoz, and M. Malkan for
valuable discussions and suggestions.  The anonymous referee is also
thanked for help improving the final paper.  This research has made
extensive use of a number of electronic databases, including the NASA
Astrophysics Data System Abstract Service, the NASA/IPAC Extragalactic
Database (NED), and the NASA/IPAC Infrared Science Archive.  These
latter two are operated by the Jet Propulsion Laboratory, California
Institute of Technology, under contract with NASA.  Primary support for
this work was provided by NASA grant AR-06380.01-A, with additional
support from the NICMOS imaging program supported by GO-07867.01-A, both
grants from the Space Telescope Science Institute, which is operated by
the Association of Universities for Research in Astronomy, Inc., under
NASA contract NAS5-26555.


%
%

%
%

\clearpage

\begin{center}
\begin{deluxetable}{llccccrrr}
\tablecolumns{9}
\tablenum{1}
\tablewidth{6.5truein}
\tablecaption{CfA Seyfert Galaxies with Archival WFPC2 Imaging\label{tbl:cfa}}
\tablehead{
\colhead{Galaxy} &
\colhead{} &
\colhead{Seyfert} &
\colhead{Exp} &
\colhead{Filter} &
\colhead{HST} &
\colhead{$d_{75}$} &
\colhead{$X_{75}$} &
\colhead{$D_{25}$}\\
\colhead{Name} &
\colhead{Other} &
\colhead{Type\tablenotemark{a}} &
\colhead{(sec)} &
\colhead{Name} &
\colhead{Prop ID} &
\colhead{(Mpc)\tablenotemark{b}} &
\colhead{(pc/\arcsec)} &
\colhead{(kpc)}
}
\startdata
Mrk 231	&	& 1	& 700	& F814W	& 5982 & 165.3 & 801 & 63.4 \\
Mrk 279	&	& 1	& 500 	& F606W	& 5479 & 124.1 & 602 & 31.4 \\
Mrk 335	&PG0003+198& 1	& 500	& F606W	& 5479 & 106.8 & 518 & 9.8  \\
Mrk 590	&NGC 863& 1	& 500	& F606W	& 5479 & 106.8 & 518 & 33.3 \\
Mrk 766	&NGC 4253& 1	& 500	& F606W	& 5479 &  50.7 & 246 & 14.1 \\
NGC 4051&	& 1	& 500	& F606W	& 5479 &  17.0 &  82 & 26.0 \\
NGC 4235&	& 1	& 500	& F606W	& 5479 &  28.7 & 139 & 34.8 \\
NGC 5940&	& 1	& 500	& F606W	& 5479 & 135.0 & 654 & 31.2 \\
NGC 7469&	& 1	& 500	& F606W	& 5479 &  66.9 & 324 & 28.8 \\
A0048+29&UGC 524&1	& 610	& F814W	& 6361 & 147.0 & 713 & 38.1 \\
Mrk 817 &	& 1.5	& 500	& F606W	& 5479 & 127.6 & 620 & 24.0 \\
Mrk 993 &	& 1.5	& 500	& F606W	& 5479 &  65.1 & 316 & 41.4 \\
NGC 3227&	& 1.5	& 500	& F606W	& 5479 &  20.6 & 100 & 32.2 \\
NGC 3516&	& 1.5	& 500	& F606W	& 5479 &  38.9 & 189 & 19.7 \\
NGC 4151&	& 1.5	& 400	& F547M	& 5433 &  20.3 &  98 & 37.3 \\
NGC 5548&	& 1.5	& 500	& F606W	& 5479 &  67.0 & 325 & 28.2 \\
NGC 6104&	& 1.5	& 500	& F606W	& 5479 & 113.5 & 550 & 27.5 \\
NGC 7603&Mrk 530& 1.5	& 500	& F606W	& 5479 & 118.4 & 574 & 53.3 \\
Mrk 334 &	& 1.8	& 500	& F606W	& 5479 &  88.4 & 429 & 25.1 \\
Mrk 471 &	& 1.8	& 500	& F606W	& 5479 & 137.3 & 666 & 35.6 \\
Mrk 744 &NGC 3786& 1.8	& 500	& F606W	& 5479 &  36.1 & 175 & 23.0 \\
UGC 12138&2237+07& 1.8	& 500	& F606W	& 5479 & 102.8 & 498 & 24.9 \\
NGC 4395\tablenotemark{c}&	& 1.9	& 600	& F814W	& 6232 &   3.6 &  17 & 13.8 \\
NGC 5033&	& 1.9	& 460	& F547M	& 5381 &  18.7 &  91 & 58.3 \\
NGC 5252&	& 1.9	& 500	& F606W	& 5479 &  90.7 & 440 & 36.4 \\
NGC 5273&	& 1.9	& 460	& F547M	& 5381 &  16.5 &  80 & 13.2 \\
NGC 5674&	& 1.9	& 500	& F606W	& 5479 &  98.1 & 476 & 31.3 \\
UM 146 &0152+06 & 1.9	& 500	& F606W	& 5479 &  71.6 & 347 & 26.2 \\
Mrk 266SW &NGC5256& 2	& 500	& F606W	& 5479 & 111.0 & 538 & 38.8 \\
Mrk 270 &NGC5283& 2	& 500	& F606W	& 5479 &  38.2 & 185 & 11.9 \\
Mrk 461 &1335+34& 2     & 560   & F606W & 8597 &  65.5 & 340 & 14.1 \\
Mrk 573 &	& 2	& 500	& F606W	& 5479 &  71.0 & 344 & 27.9 \\
NGC 1068&	& 2	& 500	& F606W	& 5479 &  14.4 &  70 & 29.7 \\
NGC 1144&	& 2	& 500	& F606W	& 5479 & 116.4 & 564 & 37.1 \\
NGC 3362&	& 2	& 500	& F606W	& 5479 & 108.7 & 527 & 44.7 \\
NGC 3982&	& 2	& 500	& F606W	& 5479 &  17.0 &  82 & 11.6 \\
NGC 4388&	& 2	& 560	& F606W	& 8597 &  16.8 &  81 & 27.5 \\
NGC 5347&	& 2	& 500	& F606W	& 5479 &  36.7 & 178 & 18.1 \\
NGC 5695&Mrk 686& 2	& 500	& F606W	& 5479 &  56.9 & 276 & 25.6 \\
NGC 5929&	& 2	& 500	& F606W	& 5479 &  38.5 & 187 & 10.7 \\
NGC 7674&Mrk 533& 2	& 500	& F606W	& 5479 & 118.5 & 575 & 38.7 \\
NGC 7682&	& 2	& 500	& F606W	& 5479 &  70.8 & 343 & 25.3 \\
UGC 6100&A1058+45& 2	& 500	& F606W	& 5479 & 117.6 & 570 & 28.5 \\
\enddata
\tablenotetext{a}{Osterbrock \& Martel (1993)}
\tablenotetext{b}{See Text, assumes $H_0=75$\,km\,sec$^{-1}$\,Mpc$^{-1}$}
\tablenotetext{c}{Classification changed from 1.0 to 1.8 by Ho et al. (1997).}
\end{deluxetable}
\end{center}

\clearpage

\begin{center}
\begin{deluxetable}{lccccccl}
\tablecolumns{8}
\tablenum{2}
\tablewidth{6.5truein}
\tablecaption{Morphological Properties\label{tbl:prop}}
\tablehead{
\colhead{Galaxy} &
\colhead{Type} &
\colhead{RC3\tablenotemark{a}} &
\colhead{D25\tablenotemark{a}} &
\colhead{R25\tablenotemark{a}} &
\colhead{D$_{kpc}$\tablenotemark{b}} &
\colhead{f$_{30}$\tablenotemark{c}} &
\colhead{Notes\tablenotemark{d}}\\
}
\startdata
Mrk 231   &  1  & .SAT5\$P& 1.12 (0.06) & 0.13 (0.05) & 63.4 & 0.38 & D,NS,Z \\
Mrk 279   &  1  & .L..... & 0.94 (0.06) & 0.23 (0.03) & 31.4 & 0.57 & D,S,NS,B?\\
Mrk 335   &  1  & .       & 0.5  (0.14) & 0    (0.12) & 9.8  & 1.58 & D,S,Z \\
Mrk 590   &  1  & .SAS1*. & 1.03 (0.03) & 0.03 (0.04) & 33.3 & 0.47 & D,NB \\
Mrk 766   &  1  & PSBS1*. & 0.98 (0.04) & 0.06 (0.04) & 14.1 & 0.52 & S,NS(GD),B,BIR \\
NGC 4051  &  1  & .SXT4.. & 1.72 (0.02) & 0.13 (0.02) & 26.0 & 0.10 & NS \\
NGC 4235  &  1  & .SAS1./ & 1.62 (0.02) & 0.65 (0.02) & 34.8 & 0.12 & INC \\
NGC 5940  &  1  & .SB.2.. & 0.9  (0.07) & 0    (0.05) & 31.2 & 0.63 & D,S,NS(GD),B,BIR \\
NGC 7469  &  1  & PSXT1.. & 1.17 (0.02) & 0.14 (0.03) & 28.8 & 0.34 & NS,NR,BIR \\
A0048+29  &  1  & PSBS3.. & 0.95 (0.07) & 0    (0.05) & 38.1 & 0.56 & D,S,NS,NR,NB,BIR \\
Mrk 817   & 1.5 & .S?.... & 0.81 (0.11) & 0    (0.06) & 24.0 & 0.77 & D,S,NS(GD),B \\
Mrk 993   & 1.5 & .S..1.. & 1.34 (0.02) & 0.52 (0.03) & 41.4 & 0.23 & INC,NS \\
NGC 3227  & 1.5 & .SXS1P. & 1.73 (0.02) & 0.17 (0.03) & 32.2 & 0.09 & NS \\
NGC 3516  & 1.5 & RLBS0*. & 1.24 (0.03) & 0.11 (0.04) & 19.7 & 0.29 & NS,BIR \\
NGC 4151  & 1.5 & PSXT2*. & 1.8  (0.02) & 0.15 (0.03) & 37.3 & 0.08 & NS \\
NGC 5548  & 1.5 & PSAS0.. & 1.16 (0.02) & 0.05 (0.03) & 28.2 & 0.35 & NS \\
NGC 6104  & 1.5 & .S?.... & 0.92 (0.07) & 0.08 (0.05) & 27.5 & 0.60 & D,S,NS(GD),B,BIR \\
NGC 7603  & 1.5 & .SAT3*P & 1.19 (0.04) & 0.18 (0.04) & 53.3 & 0.32 & D,NS \\
Mrk 334   & 1.8 &.P.....  & 0.99 (0.05) & 0.14 (0.04) & 25.1 & 0.51 & S,NS \\
Mrk 471   & 1.8 & .SB.1.. & 0.95 (0.05) & 0.19 (0.04) & 35.6 & 0.56 & D,S,NS(GD),B \\
Mrk 744   & 1.8 & .SXT1P. & 1.34 (0.02) & 0.23 (0.03) & 23.0 & 0.24 & NS \\
UGC 12138 & 1.8 & .SB.1.. & 0.92 (0.07) & 0.08 (0.05) & 24.9 & 0.60 & D,S,NS(GD),B,BIR \\
NGC 4395  & 1.9 & .SAS9*. & 2.12 (0.01) & 0.08 (0.02) & 13.8 & 0.04 & Z \\
NGC 5033  & 1.9 & .SAS5.. & 2.03 (0.01) & 0.33 (0.02) & 58.3 & 0.05 & INC,NS \\
NGC 5252  & 1.9 & .L..... & 1.14 (0.09) & 0.21 (0.05) & 36.4 & 0.36 & NS \\
NGC 5273  & 1.9 & .LAS0.. & 1.44 (0.03) & 0.04 (0.04) & 13.2 & 0.18 & NS \\
NGC 5674  & 1.9 & .SX.5.. & 1.04 (0.06) & 0.02 (0.05) & 31.3 & 0.46 & NS(GD),B,BIR\\
UM 146    & 1.9 & .SAT3.. & 1.1  (0.04) & 0.1  (0.04) & 26.2 & 0.40 & NS,BIR \\
Mrk 266SW &  2  & .P..... & 1.08 (0.06) & 0.05 (0.05) & 38.8 & 0.42 & D,INT \\
Mrk 270   &  2  & .L...?. & 1.03 (0.08) & 0.04 (0.03) & 11.9 & 0.47 & NS,NB \\
Mrk 461   &  2  & .S..... & 0.87 (0.05) & 0.13 (0.05) & 15.1 & 0.67 & NS \\
Mrk 573   &  2  & RLXT+*. & 1.13 (0.07) & 0.01 (0.04) & 27.9 & 0.37 & NS(GD),NB,BIR \\
NGC 1068  &  2  & RSAT3.. & 1.85 (0.01) & 0.07 (0.02) & 29.7 & 0.07 & NS \\
NGC 1144  &  2  & .RING.B & 1.04 (0.08) & 0.21 (0.06) & 37.1 & 0.46 & D,NS \\
NGC 3362  &  2  & .SX.5.. & 1.15 (0.05) & 0.11 (0.05) & 44.7 & 0.35 & D,NS \\
NGC 3982  &  2  & .SXR3*. & 1.37 (0.02) & 0.06 (0.03) & 11.6 & 0.21 & NS \\
NGC 4388  &  2  & .SAS3*/ & 1.75 (0.01) & 0.64 (0.02) & 27.5 & 0.09 & INC \\
NGC 5347  &  2  & PSBT2.. & 1.23 (0.04) & 0.1  (0.04) & 18.1 & 0.29 & NS(GD),B,BIR \\
NGC 5695  &  2  & .S?.... & 1.19 (0.05) & 0.15 (0.05) & 25.6 & 0.32 & NS,B,BIR \\
NGC 5929  &  2  & .S..2*P & 0.98 (0.06) & 0.03 (0.04) & 10.7 & 0.52 & S,INT,NS \\
NGC 7674  &  2  & .SAR4P. & 1.05 (0.03) & 0.04 (0.04) & 38.7 & 0.45 & D,NS(GD),B,BIR \\
NGC 7682  &  2  & .SBR2.. & 1.09 (0.04) & 0.05 (0.04) & 25.3 & 0.41 & NS,BIR \\
UGC 6100  &  2  & .S..1?. & 0.92 (0.07) & 0.18 (0.05) & 28.5 & 0.60 & D,S,NS \\
\enddata
\tablenotetext{a}{Data taken from the RC3 catalog.}
\tablenotetext{b}{Projected size of $D_{25}$ in kiloparsecs at the 
distance of the galaxy.}
\tablenotetext{c}{Fraction of $D_{25}$ contained within our 30\arcsec\  
structure maps.}
\tablenotetext{d}{Codes for the morphological classifications. D: distance 
greater than 100 Mpc (assuming $H_0 = 75$ km/s/Mpc), INC: High Inclination 
($R_{25} > 0.3$), INT: Strongly interacting system, Z: No resolved 
structure in central kiloparsec, NR: Nuclear ring, NS: Nuclear spiral, 
NS(GD): Grand-design nuclear spiral, B: Straight dust lanes indicative of a  
large-scale bar in the structure map, BIR: Bar found in the $K-$band by 
\citet{mcleod95}. }
\end{deluxetable}
\end{center}

\clearpage

%
%

\begin{figure}
\caption{
Structure maps for Mrk\,231, Mrk\,279, Mrk\,335, Mrk\,590, Mrk\,766, and
NGC\,4051. These images show the central 30\arcsec\ of each galaxy and
intensities (black to white) show the fractional residuals ($\pm10\%$)
about the original pixel-to-pixel intensity.  Dark regions show the
locations of dust obscuration, and bright regions are either locations
of enhanced stellar light (e.g., star formation regions) or
emission-line regions.  All images have been rotated to orient them
North up, East to the left.
\label{fig:stmap1} }
\end{figure}

\begin{figure}
\caption{Same as Figure~\ref{fig:stmap1} for NGC\,4235, NGC\,5940, NGC\,7469, 
A0048+29, Mrk\,817, and Mrk\,993. 
\label{fig:stmap2} }
\end{figure}

\begin{figure}
\caption{Same as Figure~\ref{fig:stmap1} for NGC\,3227, NGC\,3516, NGC\,4151, 
NGC\,5548, NGC\,6104, and NGC\,7603. 
\label{fig:stmap3} }
\end{figure}

\begin{figure}
\caption{Same as Figure~\ref{fig:stmap1} for 
Mrk\,334, Mrk\,471, Mrk\,744, 
UGC\,12138, NGC\,4395, and NGC\,5033.
\label{fig:stmap4} }
\end{figure}

\begin{figure}
\caption{Same as Figure~\ref{fig:stmap1} for 
NGC\,5252, NGC\,5273, NGC\,5674, UM\,146,
Mrk\,266SW (lower right galaxy in panel), and Mrk\,270.
\label{fig:stmap5} }
\end{figure}

\begin{figure}
\caption{Same as Figure~\ref{fig:stmap1} for 
Mrk\,461, Mrk\,573, NGC\,1144, NGC\,3362,
NGC\,3982, and NGC\,4388.
\label{fig:stmap6} }
\end{figure}

\begin{figure}
\caption{Same as Figure~\ref{fig:stmap1} for 
NGC\,5347, NGC\,5695, NGC\,5929, NGC\,7674, 
NGC\,7682, and NGC\,6100.
\label{fig:stmap7} }
\end{figure}

\begin{figure}
\caption{({\it left}) Same as Figure~\ref{fig:stmap1} for NGC\,1068 and 
the F606W image. ({\it right}) The central $10''$ of NGC\,1068 in the
F547M image, corresponding to the dashed box at left.
\label{fig:n1068} }
\end{figure}

\begin{figure}
\caption{Illustration of the ability of structure maps to recover the same 
dust and emission features seen in $(V-H)$ colormaps for NGC\,7674.  The
$(V-H)$ colormap ({\it left}) from \citet{martini99} clearly shows a
two-arm nuclear dust spiral, in addition to a PSF artifact from the
bright nuclear source in the $H$ image. The structure map ({\it right})
shows the same dust lanes as the colormap, although without the NICMOS
PSF artifacts. Each panel is 5\arcsec\ on a side.
\label{fig:vh} }
\end{figure}

\begin{figure}
\caption{Illustration of the difference between structure maps and
common unsharp masking techniques for NGC\,3516. The original image
({\it upper left}) shows some evidence of the nuclear dust spirals that
are clearly revealed with the structure map ({\it upper right}). These
spirals have much lower contrast when an unsharp masking technique is
used, either with a 2.2 pixel wide Gaussian ({\it lower left}) filter or
a 3x3 pixel Boxcar ({\it lower right}) filter. Each panel is $15''$ on a
side.
\label{fig:unsharp} }
\end{figure}

\begin{figure}
\caption{Dusty nuclear disks in (a) NGC\,5252 and (b) Mrk\,270.  The
scale is indicated in arcseconds.  Each panel shows the structure map
for the galaxy, with a ellipse illustrating the measured major axis,
ellipticity, and position angle of the dust disk.  In NGC\,5252, we draw
the axes of the galaxy disk and the ionization cone for reference.
Mrk\,270 is nearly face on and has no ionization cone, but there is
filamentary emission-line gas that appears to be illuminated material in
the farside of the nuclear disk.
\label{fig:nucdisk} }
\end{figure}

\end{document}